\documentclass[a4paper,11pt,twocolumn]{article}
\usepackage{epsfig}
\title{\bf Status of the OPERA Experiment}
\author{R.~Zimmermann${}^a$ on behalf the OPERA collaboration}

\begin{document}
\maketitle
\date{}
\begin{center}
\footnotesize${}^a$Institut f\"ur Experimentalphysik, Universit\"at Hamburg, D-22761 Hamburg, Germany\\
\end{center}

\begin{abstract}
In this article the physics motivation and the detector design of the OPERA experiment will be reviewed. The construction status of the detector, which will be situated in the CNGS beam from CERN to the Gran Sasso laboratory, will be reported. A survey on the physics performance will be given and the physics plan in 2006 will be presented.
\end{abstract}

\section{Introduction}
\label{sec:Introduction}
The CNGS project is designed to search for the $\nu_{\mu}\leftrightarrow\nu_{\tau}$ oscillation in the parameter region indicated by the SuperKamiokande, Macro and Soudan2 atmospheric neutrino analysis \cite{skbib}, \cite{macrobib}, \cite{soudanbib}. The main goal is to find $\nu_{\tau}$ appearance by direct detection of the $\tau$ from $\nu_{\tau}$ CC interactions. One will also search for the subleading $\nu_{\mu}\leftrightarrow\nu_{e}$ oscillations, which could be observed if $\theta_{13}$ is close to the present limit from CHOOZ \cite{choozbib} and PaloVerde \cite{pvbib}. In order to reach these goals a $\nu_{\mu}$ beam will be sent from CERN to Gran Sasso. In the Gran Sasso laboratory, the OPERA experiment (CNGS1) is under construction. 

At the distance of $L=732$~km between the CERN and the Gran Sasso laboratory the $\nu_{\mu}$ flux of the beam is optimized to yield a maximum number of CC $\nu_{\tau}$ interactions at Gran Sasso. With a mean beam energy of $E=17$~GeV the contamination with $\bar{\nu}_{\mu}$ is around 2~\% and with $\nu_e$ ($\bar{\nu}_e$) is less than 1~\%. The content of $\nu_{\tau}$ is negligible. The resulting $L/E$ ratio is 43~km/GeV, thus the CNGS operates in the "off peak" mode due to the high energy needed for $\tau$ production. The civil engineering has been completed, all beam parts are produced and will be installed in time. The commissioning will start in May 2006 and the first beam is expected in July 2006. Assuming a beam intensity of $4.5\times10^{19}$~pot/year and a five year run, 31000 neutrino events (CC + NC) are expected for an average target mass of 1.6~kt. At full mixing, 95 $\nu_{\tau}$ CC interactions are expected for $\Delta m^2 = 2 \times 10^{-3}$~eV$^2$ and 214 $\nu_{\tau}$ CC interactions for $\Delta m^2 = 3 \times 10^{-3}$~eV$^2$. 

The detection of the $\nu_{\tau}$'s will be done by looking for the charged $\tau$ lepton produced in $\nu_{\tau}$ CC interaction, and its decay products. The different $\tau$ decay modes can be separated into muonic (BR 18~\%), electronic (BR 18~\%) and hadronic (BR 64~\%) channels. In order to observe the decay topology, a spatial resolution at the micrometer scale is necessary. In OPERA, such a resolution is achieved by detecting the charged particle tracks in emulsion sheets interspersed with thin lead plates acting as target material. This technique, known as Emulsion Cloud Chamber (ECC), was validated for $\tau$ search in the DONUT experiment \cite{donutbib}. The basic target module is called a brick. It consists of a sandwich of 57 lead plates (1~mm thick) with in total 56 emulsion layers in between. A brick has a size of 10.2~cm$ \times 12.7$~cm with a length of 7.5~cm (10 radiation lengths) and a weight of 8.3~kg. In a separate package, two additional emulsion sheets, called changable sheets, are glued on its downstream face. The bricks are arranged in brick walls. Within a brick, the achieved spatial resolution is $\Delta x = 1\,\mu$m and the angular resolution is $\Delta \theta = 2$~mrad. With these values, the reconstruction of the $\nu$ interaction vertex and of the $\tau$ decay topology will be possible. The brick as stand alone detector also allows, to some extent, the momentum measurement of charged particles by multiple scattering, the separation of low energy $\pi$ and $\mu$ by $dE/dx$ measurements and the identification of electrons and photons by measurement of the electromagnetic shower development. To provide a $\nu$ interaction trigger and to identify the brick in which the interaction took place, the brick walls are complemented by a target and a muon spectrometer. The target tracker consist of highly segmented scintillator planes inserted between the brick walls while, downstream the target section, the magnetic spectrometer will measure the muon momenta and identify the sign of their charges. Combining the signals left in the electronic detectors by the hadronic shower and, eventually, the muon track, the brick containing the $\nu$ interaction vertex can be determined with a good efficiency. The brick is then extracted from the wall and its changable sheets are developed and scanned using fast automatic microscopes. Once the presence of tracks coming out of an interaction is confirmed, the brick is unpacked and all its emulsion sheets are developed and sent to the scanning laboratories for further analysis.
\label{sec:ConstructionStatus}
\begin{figure}
	\centering
		\includegraphics[scale=0.7]{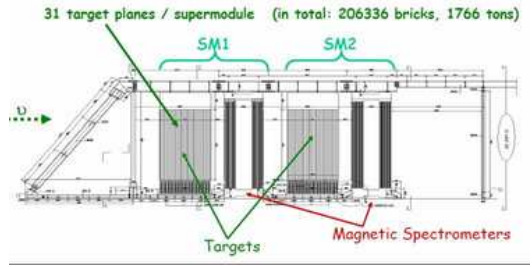}
	\caption{The Opera detector.}
	\label{fig:detopera}
\end{figure}

\section{Construction status}
The OPERA detector, shown in fig \ref{fig:detopera}, consists of two identical super-modules. Each super-module consists of a target section with 31 target planes in between the brick walls, followed by a muon spectrometer. With a total number of about 206000 bricks, an initial mass of 1.8~kt will be achieved. The target tracker covers a total area of 7000~m$^2$ and is built of 32000 scintillator strips, each 7~m long and of 25~mm$ \times $15~mm cross section. Along the strip, a wavelength shifting fiber of 1~mm diameter transmits the light signals to both ends. The readout will be done by 1000 64 channel matrix PMT's from HAMAMATSU. Since the end of the last year the data taking with cosmics has started. The target of the first super-module (yet to be filled with bricks) was installed in November 2005, the second target will be installed in June 2006. The achieved brick wall position accuracy is better than 1~mm. 

The muon spectrometer consists of the magnet instrumented with RPC's and drift tubes. The magnet is a $8 \times 8$~m$^2$ large dipole with a field of 1.55~T in upward direction on one side and in downward direction on the other side. This gives the possibility to measure the momentum twice in order to reduce the error by $\sqrt{2}$. Each magnet side consists of twelve 5~cm thick iron slabs, alternating with RPC planes. This sandwich structure allows the tracking in the magnetic field to identify the muons and to determine their momentum. In addition the precision tracker \cite{specbib} measures the muon track coordinates in the horizontal plane. It is made of 8~m long drift tubes with an outer diameter of 38~mm. These drift tubes are arranged in planes, located in front and behind the magnet as well as between the two magnet sides. The precision tracker consists of 12 drift tube planes, each covering an area of 8~m$ \times $8~m. Together with the precision tracker, the efficiency of the muon identification as well as the accuracy of the momentum measurements and sign determination can be increased; the charge misidentification is expected to be 0.1~\% - 0.3~\%. This value is important to minimize the background originating from the charmed particles produced in $\nu_{\mu}$ interactions. With the muon spectrometer a momentum resolution of $\Delta p / p \le 0.25$ for all muon momenta $p$ up to a maximum of $p=25$~GeV/c can be achieved. To reduce the number of ghost tracks two planes of glass RPC's (XPC's), each consisting of two 45~$^{\circ}$ crossed planes, are installed in front of the magnets.

\begin{figure}
	\centering
		\includegraphics[scale=0.7]{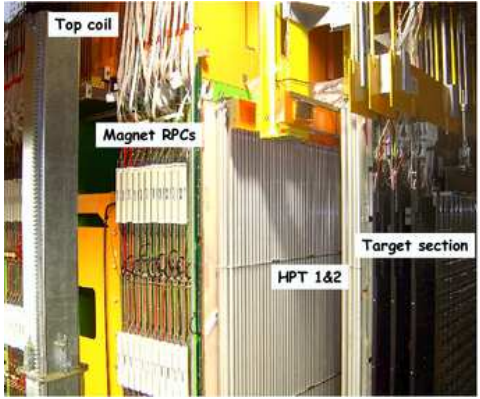}
		\vspace{0.6cm}
	\caption{The muon spectrometer.}
	\label{fig:elecspectrometer}
\end{figure}
The construction status at the end of 2005 is shown in fig \ref{fig:elecspectrometer}. The brick supporting structure and all the tracker planes of the first super-module are installed. The XPC's and three of the high precision tracker planes of the first super-module are installed. The magnets, including the RPC's for the whole detector and the mechanical structure are completed.
\begin{figure}
	\centering
		\includegraphics[scale=0.8]{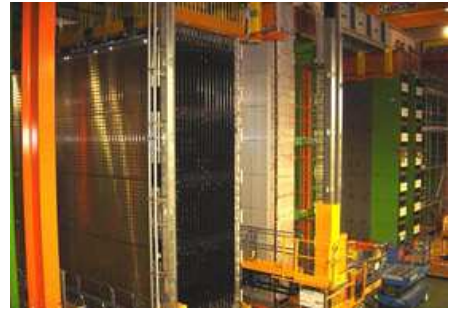}
	\caption{The total detector.}
	\label{fig:totaldet}
\end{figure}
To handle the data flow a new DAQ system was developed. It will use a Gigabit network consisting of 1200 nodes. To match the data of the different subdetectors a event time stamp will be used. This will be delivered by a clock distribution system using the GPS\footnote{Global Positioning System}. The heart of the DAQ is the Mezzanine chip. It contains the CPU, the memory, the clock receiver for the time stamp and the ethernet connections to the other components. All the components of the DAQ system are available and the final version of the Mezzanine is under test.

The production of the bricks will be done by the brick assembling machine (BAM). It consists of robots for the mechanical packing of the bricks. In total 23 million lead plates and emulsion layers have to be assembled to 206000 bricks. The required production speed will be 2 bricks per minute. The final system was successfully tested and is ready for installation at the Gran Sasso laboratory.

Inside the detector the produced bricks will be handled by the brick manipulator system (BMS). Two large robots, each one operating at one side of the detector, consist of a drum for the brick transfer and a brick storage carousel. A pushing arm will be used to insert the bricks in any given row. The extraction of a brick in the region of interest indicated by the electronic detectors will be done by a vacuum sucker. The first BMS robot was installed successfully end of 2005 and is in the commissioning phase. A picture of the construction status end of 2005 is shown in fig \ref{fig:totaldet}.

\section{Physics performance}
\label{sec:PhysicsPerformance}
The detection efficiency of a tau decay has been studied by detailed Monte Carlo simulations. One may distinguish two cases: in so-called short decays, the tau decays in the lead plate in which it is produced and the only signature is a non-zero impact parameter of the decay products with respect to the primary neutrino vertex. This vertex can only be determined reliably for multi-prong deeply inelastic scatterings (DIS). On the contrary, for "long" decays, the tau track direction can be directly measured and the kink angle of the charged decay products accurately determined, both for DIS and quasi elastic (QE) neutrino interactions. The resulting $\tau$ detection efficiencies for the detected channels, weighted with their branching ratios, are listed in table \ref{tab:TauDetectionEfficiencies}.
\begin{table}
	\centering
		\begin{tabular}{|l|c|c|c|c|}
		\hline
		\small
			& \footnotesize DIS long &\footnotesize QE long &\footnotesize DIS short &\footnotesize overall\\\hline
			$\tau \rightarrow$ e & 2.7 & 2.3 & 1.3 & 3.4\\
			$\tau \rightarrow \mu $& 2.4 & 2.5 & 0.7 & 2.8\\
			$\tau \rightarrow$ h & 2.8 & 3.5 & - & 2.9\\
			total & 8.0 & 8.3 & 1.3 & 9.1\\\hline
		\end{tabular}
	\caption{The $\tau$ detection efficiencies in \% for the different decay channels of the  $\tau$ lepton.}
	\label{tab:TauDetectionEfficiencies}
\end{table}

\begin{table*}
	\centering
		\begin{tabular}{|c|c|c|c|c|}
		\hline
		$\Delta m^2$	& \footnotesize $1.9 \times 10^{-3}$~eV$^2$&\footnotesize $2.4 \times 10^{-3}$~eV$^2$ &\footnotesize $3.0 \times 10^{-3}$~eV$^2$ &\footnotesize BKGD\\\hline
			1.8~kt fiducial & 6.6 (10) & 10.5 (15.8) & 16.4 (24.6) & 0.7 (1.1)\\
		  \footnotesize + improved brick finding + 3 prong decay & 8.0 (12.1) & 12.8 (19.2) & 19.9 (29.9) & 1.0 (1.5)\\
		+ BKGD reduction & 8.0 (12.1) & 12.8 (19.2) & 19.9 (29.9) & 0.8 (1.2)\\\hline
		\end{tabular}
	\caption{The expected number of $\tau$ events. The numbers in brackets are for a CNGS beam upgrade (factor 1.5 increase in intensity).}
	\label{tab:NumberTauEvents}
\end{table*}

\begin{figure}
	\centering
		\includegraphics[scale=0.9]{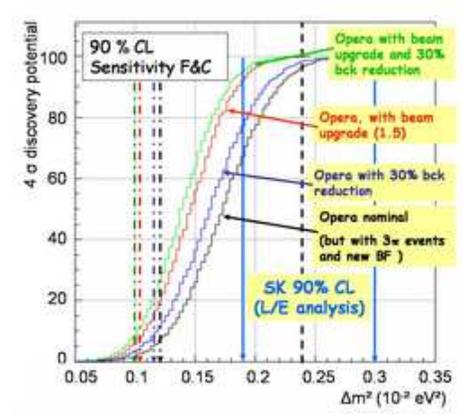}
		\caption{The OPERA sensitivity (vertical lines) and the 4$\sigma$ discovery probability (solid lines).}
	\label{fig:phpot}
\end{figure}

The first OPERA run is scheduled for the second half of 2006. After the completion and the commissioning of the CNGS beam, a low intensity run with integrated $0.3 \times 10^{19}$~pot will be performed. After the super-module 1 is partially filled with bricks, there will be a normal intensity run with a total of $\approx 4 \times 10^{19}$~pot. This physics run is important for the validation and monitoring of the CNGS beam, which will be done by checking the interaction rates, the energy distribution and by analyzing the muon charge to determine the content of $\bar{\nu}_{\mu}$. The collected sample of interactions in the bricks will be sufficient to check the full analysis procedure and control the vertex finding efficiencies estimated by simulation. Furthermore the beam induced background sources can be measured and the kinematical analysis can be tuned. In table \ref{tab:NumberTauEvents} the expected number of $\tau$ events is listed for a nominal beam intensity of $4.5 \times 10^{19}$~pot/year. The values in brackets apply for an increase in beam intensity by a factor of 1.5. The resulting sensitivity and the 4$\sigma$ discovery probability, the probability that one can claim to have seen a signal larger than a 4$\sigma$ fluctuation of the background, are shown in fig \ref{fig:phpot}. This probability is described by the solid lines dependent on the value of $\Delta m^2$. The OPERA nominal run with an average target mass of 1.6~kt includes an improved brick finding and the $3\pi$ hadronic decay channel. The probability can be increased by reducing the background by 30~\% an by increasing the beam intensity by a factor of 1.5. In the case that no $\nu_{\tau}$ appearance will be seen, an exclusion region (90~\% CL) can be given. These sensitivities (vertical lines) are computed using the feldman and cousins method. In the plot the values correspond to full mixing. The $3\pi$ decay channel is included.

As mentioned before OPERA will also search for $\nu_{\mu}\leftrightarrow\nu_{e}$ oscillations. If $\theta_{13}$ is close to the CHOOZ limit ($\sin^2 2\theta_{13} \ge 0.14$ for $\Delta m^2 = 2.5 \times 10^{-3}$~eV$^2$) \cite{choozbib}, OPERA has the potential to observe the appearance of $\nu_{e}$. In case no $\nu_{e}$ are observed and assuming $\Delta m^2 = 2.5 \times 10^{-3}$~eV$^2$, OPERA will be able to set a limit $\sin^2 2\theta_{13}<0.06$ (90\%~CL) \cite{t13bib}.

\section{Conclusion}
\label{sec:Conclusion}
The OPERA experiment was designed to search for the $\nu_{\mu}\leftrightarrow\nu_{\tau}$ oscillation. A search for $\nu_{\mu}\leftrightarrow\nu_e$ oscillation will also be performed, in order to measure the angle $\theta_{13}$. The project will start operating in July 2006. The discovery of the $\nu_{\tau}$ appearance in $\nu_{\mu}\leftrightarrow\nu_{\tau}$ oscillation is an important part of the neutrino oscillation puzzle. OPERA will also prepare the way for future neutrino appearance experiments. 

\end{document}